\documentclass[useAMS,usenatbib]{mn2e}
\usepackage{epsfig}
\usepackage{amsmath}

\newcommand{\be}{\begin{equation}}
\newcommand{\beq}{\begin{equation}}
\newcommand{\ba}{\begin{eqnarray}}
\newcommand{\ee}{\end{equation}}
\newcommand{\eeq}{\end{equation}}
\newcommand{\ea}{\end{eqnarray}}

\newcommand{\hs}{\hspace{1mm}}

\newcommand{\dPsidL}{$\partial \log \Psi/\partial \log L_B$}
\newcommand{\apj}{ApJ}
\newcommand{\aap}{A\&A}
\newcommand{\apjl}{ApJL}
\newcommand{\mnras}{MNRAS}
\newcommand{\aj}{AJ}
\newcommand{\apjs}{ApJS}
\newcommand{\nat}{{\it Nature}}
\newcommand{\pasj}{PASJ}

\def\lsim{~\rlap{$<$}{\lower 1.0ex\hbox{$\sim$}}}

\def\gsim{~\rlap{$>$}{\lower 1.0ex\hbox{$\sim$}}}

\title[Ly$\alpha$ Constraints on Very Low Luminosity AGN]{Ly$\alpha$ Constraints on Very Low Luminosity AGN}

\author[Mark Dijkstra and J. Stuart B. Wyithe]{Mark Dijkstra\thanks{E-mail:dijkstra@physics.unimelb.edu.au} and J. Stuart B. Wyithe\\
School of Physics, University of Melbourne, Parkville, Victoria, 3010, Australia}

\begin{document}

\date{\today}
\pagerange{\pageref{firstpage}--\pageref{lastpage}} \pubyear{2006}

\maketitle

\label{firstpage}
\begin{abstract}
Recent surveys have detected Ly$\alpha$ emission from
$z=4.5-6.5$ at luminosities as low as  $10^{41}$ erg s$^{-1}$.
 There is good evidence that low numbers of AGN are among observed
faint Ly$\alpha$ emitters. Combining these observations with an
empirical relation between the intrinsic Ly$\alpha$ and B-band
luminosities of AGN, we obtain an upper limit on the number
density of AGN with absolute magnitudes $M_B \in [-16,-19]$ 
at $z=4.5-6.5$. These AGN are up to two orders of magnitude fainter than those discovered in the Chandra Deep Field, resulting in the faintest observational
constraints to date at these redshifts. At $z=4.5$, the powerlaw slope of the very faint end of the luminosity function of AGN is shallower than the slope observed at lower redshifts, $\beta_l <1.6$, at the 98\% confidence level. In fact, we find marginal evidence that the luminosity function rises with luminosity, corresponding to a powerlaw slope $\beta_l <0 $, at magnitudes fainter than $M_B\sim -20$ (75\% confidence level). These results suggest either that accretion onto lower mass black holes is less efficient than onto their more massive counterparts, or that the number of black holes powering AGN with $M_B\gsim-20$ is lower than expected from the $M_{\rm BH}-\sigma$ relation by one-two orders of magnitude. Extrapolating from reverberation-mapping studies suggests that these black holes would have $M_{\rm BH}=10^6-10^7 M_{\odot}$. To facilitate the identification of AGN among observed Ly$\alpha$ emitters, we derive observational properties of faint AGN in the Ly$\alpha$ line, as well as in the X-ray and optical bands.
\end{abstract}

 
\section{Introduction}
\label{sec:intro}
The quasar luminosity function (QLF) describes the space density of
Active Galactic Nuclei (AGN) as a function of luminosity and
redshift. The QLF encodes information on quantities like the black hole
number density per unit mass, and the gas accretion efficiency. It
therefore constrains physical models of AGN and of super massive black
hole formation. The optical (B-band) QLF at $z \lsim 4$ has been
determined accurately for luminosities\footnote{One solar B-Band
luminosity, denoted by $L_{B,\odot}$, corresponds to $4 \times
10^{32}$ ergs s$^{-1}$.} exceeding $L_B \gsim 10^{11}L_{B,\odot}$
from the 2dF quasar survey (which corresponds to absolute magnitudes 
$M_B<-22$, Boyle et al. 2000; Croom et al. 2004). At higher 
redshifts the optical luminosity function of luminous quasars has been
determined from quasars in the Sloan Digital Sky Survey 
\citep{Fan01} at $M_B\sim-27$. In addition, deep {\it Chandra} and {\it XMM}
imaging has constrained the X-ray QLF at X-Ray
luminosities as low as $L_X=10^{42}-10^{44}$ ergs s$^{-1}$
\citep{Barger03,Cowie03,Hasinger05}. Following \citet{Haiman98},
\citet{Wyithe03} have used the X-Ray QLF to
constrain the B-band QLF down to fainter optical luminosities ($M_B
\sim -22$) at $z \gsim 4$.  At these redshifts, no observational
constraints exist on the optical QLF at fainter luminosities.
 However, the details of the existence and evolution of low
 luminosity AGN is crucial for our
understanding of the growth of low mass black holes, and of their role
in the formation of super massive black holes.

In this paper we demonstrate how existing Ly$\alpha$
surveys may be used to constrain the B-Band QLF at absolute magnitudes as low as 
$M_B =-16$. Existing wide-field narrow-band surveys \citep[e.g.][]{Rhoads00} are optimised to detect Ly$\alpha$ line emission from high-redshift galaxies, and in order to maximise their detection rate, deeply image fields as large as 1 deg$^{2}$ on the sky \citep[e.g.][]{Taniguchi05}. The combination of wide field and deep images, together with concentration on a strong emission line, rather than continuum, allows these wide-field narrow-band surveys to put stringent constraints on the number density of AGN with $M_B\sim -19$. Constraints on even fainter AGN are derived from deep spectroscopic surveys of regions around intermediate redshift clusters of galaxies, which offer an ultra-deep view into the high redshift universe (z=4.5-6.7) through strong gravitational lensing (with magnification factors of 10-1000, Santos et al, 2004). 

Throughout this paper, the terms AGN and quasar are interchangeable and
 refer to broad lined active galactic nuclei, also known as 'Type I' AGN. 
The outline of this paper is as follows: In
\S~\ref{sec:lum} we relate the Ly$\alpha$ luminosity of a quasar to
its B-band luminosity.  In \S~\ref{sec:lya} we summarise constraints
on the number density of Ly$\alpha$ emitters at high redshift, and
discuss the abundance of quasars among these sources. We show how this
existing data constrains the faint end of the quasar luminosity
function.  In \S~\ref{sec:phys} we calculate the observable Ly$\alpha$
properties of AGN and use these to physically interpret our
results. We also model the appearance of faint AGN in the optical and
X-ray bands.  Finally, in \S~\ref{sec:discuss} we discuss the
possible cosmological implications of our work, before presenting our
conclusions in \S~\ref{sec:conclusion}. We use the {\it WMAP}
cosmological parameters: $(\Omega_m, \Omega_{\Lambda}, \Omega_b, h,
Y_{\rm He})$ =$(0.3,0.7,0.044,0.7,0.24)$ \citep{Spergel03} throughout
the paper.

\section{The Relation between Ly$\alpha$ and B-band Luminosities in AGN.}
\label{sec:lum}

There is a very simple relation between the Ly$\alpha$ and B-band
luminosities of a quasar. The continuum flux density of a quasar may be denoted either 
by $F_E$ ( in units of erg s$^{-1}$ cm$^{-2}$ erg $^{-1}$) or
$F_\lambda$ ( in units of erg s$^{-1}$ cm$^{-2}$ \AA
$^{-1}$). These quantities are related by the equality $\lambda F_{\lambda}=EF_E$, where $E=h_p c/\lambda$ is the photon energy and $h_p$ is Planck's constant. Assuming the Ly$\alpha$ line has an equivalent width EW, the
Ly$\alpha$ luminosity of a quasar becomes
\begin{equation}
L_{{\rm Ly}\alpha}={\rm EW} \hs F_{\lambda}|_{\lambda=1200\AA}=\Big{(}\frac{{\rm EW}}{1200 \hs \AA}\Big{)}[\lambda F_{\lambda}]_{\lambda=1200 \AA}.
\label{eq:llya1}
\end{equation} The B-band luminosity can be estimated from

\begin{equation}
L_B={\rm BW} \hs F_{\lambda}|_{\lambda=4400\AA}=\Big{(}\frac{800}{4400}\Big{)}[\lambda F_{\lambda}]_{\lambda=4400 \AA},
\end{equation} where we assumed the B-band filter to be centered on $\lambda=4400$ \AA \hs with a width of BW$=800$ \AA. The ratio of $L_{{\rm Ly}\alpha}$ and $L_B$ becomes

\begin{equation}
\frac{L_{{\rm Ly}\alpha}}{L_B}=5.5\Big{(}\frac{{\rm EW}}{1200\hs \AA}\Big{)}\frac{[E F_{E}]_{\lambda=1200 \AA}}{[E F_E]_{\lambda=4400 \AA}}.
\label{eq:lyalb}
\end{equation} 
\citet{Sazonov04} have computed the characteristic angle-averaged, 
broad-band spectral energy distribution of the typical quasar. They
show that in the range $E=1-10$ eV, $EF_E \propto E^{\gamma}$, with
$\gamma=0.4$. The exact value of $\gamma$ varies between individual
objects. \citet{Fan01} found that  at $z\sim 4.5$ the spread in 
$\gamma$ could be represented by a Gaussian with $\sigma_\gamma \sim 0.3$ in their sample of 39 quasars.

The intrinsic Ly$\alpha$ equivalent width for AGN is
uncertain. \citet{Charlot93} showed that AGN which are completely
surrounded by neutral hydrogen gas have Ly$\alpha$ EWs of $827
\alpha^{-1}(3/4)^{\alpha} $ \AA, where $\alpha$ is the spectral index
blueward of the Ly$\alpha$ line. According to the template of
\citet{Sazonov04}, $\alpha=1.7$, which yields EW $\sim 300$
\AA. However, the observed Ly$\alpha$ EWs of bright AGN are typically
in the range 50-150 \AA, which could be attributed to dust attenuation
of the Ly$\alpha$ line \citep[see][and references therein]{Charlot93}
and/or scattering in the IGM (see \S~\ref{sec:lyaagn}). \citet{Fan01}
represent the distribution of observed EW by a Gaussian centered
on EW$=50$ \AA, with $\sigma_{\rm EW}=14$ \AA.
\footnote{\citet{Fan01} actually measure the sum 
of the Ly$\alpha$ and NV equivalent width to be EW$=70\pm 20$ \AA.
The Ly$\alpha$ line makes up $\sim 70\%$ of the total EW.}
In \S~\ref{sec:lyaagn} and Appendix~\ref{app:trans} we have 
argued in detail that $\sim 50\%$ of all Ly$\alpha$ photons
emitted by AGN is transmitted through the IGM (at $z < 6$;
at $z=6.5$ this number is $\sim 30-40 \%$).
If half of the emitted Ly$\alpha$ photons are scattered in
the IGM, then the intrinsic distribution of EW should be centered on
EW$=100$ \AA with $\sigma_{\rm EW}=30$ \AA. 
The distribution of the ratio $L_{{\rm Ly}\alpha}/L_B$ was obtained
via Monte-Carlo, under the assumption that: 1) the slope $\gamma$
follows a Gaussian distribution with $\gamma=0.4 \pm 0.3$; and 2) the
Ly$\alpha$ equivalent width follows a Gaussian distribution with
EW=$100 \pm 30$ \AA, which is truncated at EW$=0$ and 600 \AA. The
ratio $L_{{\rm Ly}\alpha}/L_B$ and it's $95\%$ confidence interval are
found to be

\begin{equation}
\frac{L_{{\rm Ly}\alpha}}{L_B}=0.7^{+1.2}_{-0.4}.
\label{eq:key}
\end{equation}  
\noindent Thus, within the uncertainties we may approximate {\it the
Ly$\alpha$ luminosity of an AGN as being equal to its B-band
luminosity}. 

It is important to stress that the results that are derived 
in this paper assume that this ratio does not change toward
 lower AGN luminosity. This effectively means that we assume
the AGN's Ly$\alpha$ EW is independent of luminosity. However,
observations of both high and low redshift quasars and Seyfert I galaxies
revealed that the EW of the CIV 1549 line increased toward lower
luminosities. The increase of the EW in certain emission lines
in spectra of AGN toward lower AGN luminosities is known
 as the 'Baldwin-Effect'. This effect, however, has not been
observed for the Ly$\alpha$ line \citep{Baldwin}. We point
out that even if the Baldwin-Effect were applicable, then this
would increase the ratio $L_{{\rm Ly}\alpha}/L_B$. For a fixed
Ly$\alpha$ luminosity, we would then be able to probe AGN
with fainter $L_B$, which would strengthen the main 
results presented in this paper.

\section{Constraining the Faint end of QLF using Ly$\alpha$ Surveys.}
\label{sec:lya}

\subsection{The Number Density of AGN at $z=4.5-6.5$.}
\label{sec:numden}

Several Ly$\alpha$ surveys have derived the number density, $n_{{\rm
Ly}\alpha}$, of Ly$\alpha$ emitters brighter than a minimum detectable
Ly$\alpha$ luminosity, $L_{{\rm Ly}\alpha{\rm,c}}$.
Examples include surveys at $z=4.5$ \citep{Dawson04}, $z=5.7$ \citep[e.g.][]{Hu04,Ouchi05,Shima06}, $z=6.5$ \citep{Taniguchi05,Kashi06} and $z>4.5$ \citep{Santos04lens}. Results from surveys that were used for this paper are summarised in columns 1-4 in Table~\ref{table:numden}.
\begin{table*}
\caption{Ly$\alpha$ Survey's Constraints on $n_{{\rm Ly}\alpha}$ and the corresponding upper limits on $\Psi(L_{\rm B,min},z)$.}
\label{table:numden}
\begin{tabular}{|lcccccc|}
\hline
z & $f_{\rm min}$ & $L_{{\rm Ly}\alpha{\rm,c}}$ & $n_{{\rm Ly}\alpha}$ & $n_{\rm AGN}$ & $L_{\rm B,min}$ & $\Psi(L_{\rm B,min},z)$ \\
&($10^{-18}$ erg s$^{-1}$ cm$^{-2}$) &($10^{42}$ erg s$^{-1}$) & ($10^{-4}$ Mpc$^{-3}$) & ($10^{-6}$ Mpc$^{-3}$) & $10^9L_{B,\odot} $&($10^{-6}$ Gpc$^{-3}$ L$_B^{-1}$)\\
\hline
\hline
$4.5^{1}$ & 16 & 6& 1.6 & $<2$ & 20 & $<0.08$ ($\beta=1.1$)\\
$5.7^2$ & 6& 4& 3.9& $< 58$ & 14 & $<2.6$ ($\beta=1.6$)\\
$6.6^{3}$ & 4.1& 6& 1.2& $< 35$ & 20 & $< 1.1$ ($\beta=1.6$)\\
$>4.5^{4}$& 0.1& 0.06 & $10^{2.0}$& $< 10^{4.8}$ & 0.7 & $< 10^{4.9}$ ($\beta=1.6$)\\
$>4.5^{4}$ & 0.3& 0.2& $10^{1.5}$& $< 10^{4.1}$& 2.2 & $< 10^{3.7}$ ($\beta=1.6$) \\
\hline
\multicolumn{7}{l}{\footnotesize{(1) Data from \citet{Dawson04}. Note that
the actual detection limit is quoted as $\sim 2 \times 10^{-17}$ erg s$^{-1}$ cm$^{-2}$. We have slightly}} \\
\multicolumn{7}{l}{\footnotesize{revised their detection limit downwards, as $27\%$ of their spectroscopically confirmed LAEs have a flux $<2.0 \times 10^{-17}$ erg s$^{-1}$ cm$^{-2}$,}}\\
\multicolumn{7}{l}{\footnotesize{with a mean flux of $\sim 1.1\times 10^{-17}$ erg s$^{-1}$ cm$^{-2}$. As argued in \S~\ref{sec:numden}, the $5-\sigma$ detection threshold for AGN lies higher}}\\
\multicolumn{7}{l}{\footnotesize{by a factor of 1.4 at $\sim 1.6\times 10^{-17}$ erg s$^{-1}$ cm$^{-2}$.}}\\
\multicolumn{7}{l}{\footnotesize{(2) Data from \citet{Ouchi05} }} \\
\multicolumn{7}{l}{\footnotesize{(3) Data from \citet{Taniguchi05} and \citet{Kashi06}}} \\
\multicolumn{7}{l}{\footnotesize{(4) From the deep spectroscopic survey of gravitationally lensed regions 
by \citet{Santos04lens}. We have increased $L_{\rm B,min}$ by a}} \\
\multicolumn{7}{l}{\footnotesize{factor of $\sqrt{10}$ (relative to that given by Eq~\ref{eq:key}) to account for the reduced sensitivity to detect the AGN's broad Ly$\alpha$ emission}}\\
\multicolumn{7}{l}{\footnotesize{lines in this spectroscopic survey (see the main text for more discussion on this).}} \\
\end{tabular}
\end{table*}

Since the detected Ly$\alpha$ emitters are generally identified as
galaxies, the number density of quasars with $L_{{\rm Ly}\alpha}>$
$L_{{\rm Ly}\alpha{\rm,c}}$ must be smaller than $n_{{\rm
Ly}\alpha}$. \citet{Dawson04} find that no AGN are among the
Ly$\alpha$ emitters in their sample in a total comoving volume of $1.5
\times 10^6$ Mpc$^3$. This finding was based on the absence of broad line emitters in both their narrow band survey\footnote{The survey volume of \citet{Dawson04} is determined by 5 overlapping filters each of width $80$ \AA, yielding a total wavelength coverage of $240$ \AA. As will be demonstrated in \S~\ref{sec:lyaagn}, the observed FWHM of the Ly$\alpha$ line of AGN at this redshift is typically $\sim 60-70$ \AA, which implies that AGN would have been detected in more than two filters. From narrow-band imaging alone, \citet{Dawson04} find no evidence for the presence of broad line emitters. If we represent the filter transmission curves with top hats of width $80 \AA$, and an AGN's Ly$\alpha$ emission line with a Gaussian with a FWHM of $65\AA$, then on average $\sim 70\%$, $60\%$ and $40\%$ of the AGN's total Ly$\alpha$ flux is detected by three adjacent filters. This implies that 1) for an AGN to be detected in 1 filter at the 5-$\sigma$ detection threshold, requires its Ly$\alpha$ flux to be 1/0.7$\sim 1.4$ times higher than the surveys detection threshold, and 2) for an AGN to be detected in 3 adjacent filters requires its peak flux to lie within the inner three narrow band filters. This reduces the survey volume in which these AGN can be identified by a factor of $(240-2\times 40)/240=2/3$.} and in follow-up high resolution spectra of 18 confirmed $z=4.5$ Ly$\alpha$ emitters. Furthermore, these objects
 lacked high--ionization state UV emission lines, symptomatic of AGN activity. Additionally, deep X-Ray observations of 101 Ly$\alpha$
emitters by Wang et al. (2004, also see Malhotra et al. 2003),
 revealed no X-ray emission from any
individual source at a $3-\sigma$ detection limit of $F_{0.5-10.0{\rm
keV}}=6.6 \times 10^{-16}$ erg s$^{-1}$ cm$^{-2}$ (but
this may not be that surprising as we discuss in \S~\ref{sec:Xray}).
The absence of AGN in the survey volume yields an upper limit on 
the number of AGN, $N_{\rm AGN} \leq 3$ at the $95\%$ confidence
level.\footnote{This is because for a Poisson distribution
$P(n,\mu)=e^{-\mu} \mu^n/n!$ with an expected number of events
$\mu=3.0$, the probability of having more than 0 events occur is
$95\%$.} 

Similarly, \citet{Ouchi05} found 515 objects within their survey volume of $1.3 \times 10^6$ Mpc$^3$ that are probable $z=5.7$ Ly$\alpha$ emitters.
Follow-up spectroscopy of 19 emitters revealed that their spectra are
too narrow ($v_{\rm FWHM} \lsim 500$ km s$^{-1}$) to be AGN. If 
the sample of 515 detections contains a fraction of AGN, $f_{\rm AGN}\lsim 0.15$, then the probability that a random sample of 19 contains
no AGN is $(1-f_{\rm AGN})^{19}\gsim 0.05$. This yields a $2-\sigma$ upper limit on the number of AGN, $N_{\rm AGN} \lsim 515f_{\rm AGN}=75$.

\citet{Taniguchi05} and \citet{Kashi06} found 53 candidate $z=6.5$ Ly$\alpha$ emitters within their survey volume of $2 \times 10^5$ Mpc$^3$ centered on the Subaru Deep Field. Follow-up spectroscopy of 22 candidates revealed a combined 17 objects that are likely to be $z=6.5$ Ly$\alpha$ emitters. The line widths of their objects lie in the range 180-480 km s$^{-1}$. This, in combination with 
the lack of NV $\lambda=1640$ \AA \hs emission, strongly suggests no AGN are in their sample. Following the discussion above, if the sample of 53 candidate Ly$\alpha$ emitters contains a fraction of AGN, $f_{\rm AGN}\lsim 0.12$, then the probability that a random sample of 22 contains no AGN is $(1-f_{\rm AGN})^{22}\gsim 0.05$. This yields a $2-\sigma$ upper limit on the number of AGN, $N_{\rm AGN} \lsim 53f_{\rm AGN}=7$.

In their deep, blind, spectroscopic survey of regions that utilized strong-lensing magnification by $z=0.2$ clusters of galaxies, \citet{Santos04lens} found 3-5 Ly$\alpha$ emitting objects at $z>4.5$ at unlensed flux levels as low as $\sim 3 \times 10^{-19}$ erg s$^{-1}$ cm$^{-2}$. The line widths of these sources are not reported, but \citet{Ellis01} found the line width of one $z=5.6$ Ly$\alpha$ emitter to be $\lsim 100$ km s$^{-1}$, and is therefore unlikely an AGN. However, to be conservative, our upper limits are derived under the assumption that all these Ly$\alpha$ emitters could be AGN.

For each survey, we obtain upper limits on the number density of AGN 
from $n_{\rm AGN}\lsim N_{\rm AGN} V_{\rm survey}^{-1}$. The detection 
limit of each survey, $f_{\rm min}$, is converted to its minimum
 Ly$\alpha$ detectable Ly$\alpha$ luminosity, $L_{{\rm Ly}\alpha{\rm, c}}$, using $L_{{\rm Ly}\alpha{\rm, c}}$= $4\pi d_L^2(z)\hs f_{\rm lim}/(\mathcal{T})$. Here, $d_L(z)$ is the luminosity distance to redshift z and $\mathcal{T}$ is the mean transmitted Ly$\alpha$ flux from the AGN through the IGM. As will be discussed in \S~\ref{sec:lyaagn}, we assume that for AGN $\mathcal{T}=0.5$ when $z < 6$ and $\mathcal{T}=0.3$ when $z = 6.5$. Values of $f_{\rm lim}$ are given in Table~\ref{table:numden}. For the survey performed by \citet{Santos04lens}, the calculation of $n_{\rm AGN}$ is far more complicated, since both the detection limit and survey volume are functions of the position on the sky and of redshift. Fortunately, \citet{Santos04lens} fully account for this and provide number densities of Ly$\alpha$ emitters with $L>L_{\rm crit}$, for several values of $L_{\rm crit}$. We obtain upper limits on $n_{\rm QSO}$ from the upper boundary of the 95\% confidence levels on their quoted $n_{{\rm Ly}\alpha}(L>L_{\rm crit})$ for $\log[L_{\rm crit}]=40.5$ and $41$ (their Figure 12. Our $L_{{\rm Ly}\alpha{\rm, c}}=2L_{\rm crit}$ because we account for 50\% loss of flux in the IGM). Because the Ly$\alpha$ emission lines of AGN are expected to be much broader than those of galaxies (typically by a factor of $\sim 10$, see \S~\ref{sec:lyaagn}), these are more difficult to detect in spectroscopic surveys, in which the AGN's Ly$\alpha$ flux would be spread out over $\sim 10$ times as many frequency bins. This would increase the total noise by a factor of $\sim \sqrt(10)$. To represent the decreased sensitivity to Ly$\alpha$ emission lines emitted by AGN, the value of $L_{\rm B,min}$ shown in Table~\ref{table:numden} was obtained from $L_{\rm B,min}=\sqrt(10) L_{{\rm Ly}\alpha{\rm ,c}}/0.7$.

A summary of the minimum detectable Ly$\alpha$ luminosity, $L_{{\rm Ly}\alpha{\rm, c}}$ and our constraints on the number densities of AGN at
$z=4.5$, $5.7$ and $6.5$ is given in Table~\ref{table:numden}. By converting $L_{{\rm Ly}\alpha{\rm,c}}$ to a minimum B-band luminosity $L_{\rm B,min}$ (Eq.~\ref{eq:key}), we are able to constrain the B-band QLF. This process is described in the following subsections. 
\subsection{Using the Number Density of AGN at $z=4.5-6.5$ to Constrain the QLF.}
\label{sec:psi}

The luminosity function $\Psi(L_B,z)dL_B$ is defined as the number of
quasars per unit comoving volume having rest-frame B-band luminosities
in the range between $L_B$ and $L_B + dL_B$ at redshift z. The number
density of AGN with $L_B$ $>$ $L_{\rm B,min}$ is then given by
\begin{equation}
n_{\rm AGN}(L_{B}>L_{\rm B,min};z)=\int_{L_{\rm B,min}}^{\infty}\Psi(L_B,z)dL_B
\label{eq:nqso}
\end{equation} 
To constrain the luminosity function at $L_{\rm B,min}$ we must know
the functional form of $\Psi(L_B,z)$, which is not constrained
observationally at the redshift range of interest, $z \in
[4.5,6.5]$. We first constrain $\Psi(L_B,z)$ at $L_{\rm B,min}$ 
at $z > 5$ under the assumption that the faint end of the quasar 
luminosity function observed at lower redshifts
applies here as well. This is followed by a self-consistent 
constraint on $\Psi(L_B,z)$ and its slope at $z=4.5$.
\subsubsection{Constraints on $\Psi(L_B,z)$ at $z>5$.}
The following double power law provides a good representation of the observed quasar luminosity function at $z \leq 3$ \citep[][]{Boyle00}
\begin{equation}
\Psi(L_B,z)=\frac{\Psi_*/L_{*}(z)}{\big{(}\frac{L_B}{L*(z)} \big{)}^{\beta_h}+\big{(} \frac{L_B}{L*(z)}\big{)}^{\beta_l}}.
\label{eq:psiobs}
\end{equation} 
The slope at the bright and faint end of the luminosity function is
$\beta_h=3.52$ and $\beta_l=1.66$, respectively. All redshift
dependence lies in the transition luminosity $L_*(z)$. If the same parameterization holds at $z> 5$ and $L_{\rm B,min} \ll L_*(z)$,
then the integral in Eq~(\ref{eq:nqso}) is dominated by the
faint end of the luminosity function and the $\beta_h-$term may be
omitted. Eq~(\ref{eq:nqso}) then becomes
\begin{equation}
n_{\rm AGN}(L_{B}>L_{\rm B,min};z)=\frac{L_{\rm B,min}}{\beta_l-1}\Psi(L_{\rm B,min},z),
\label{eq:psinum}
\end{equation} 
which shows that the upper limit on $n_{\rm AGN}(L_{B}>L_{\rm
B,min};z)$ translates directly to an upper limit on $\Psi(L_{\rm
B,min},z)$. These constraints are summarised in the last column
of Table~\ref{table:numden} (at $z=5.7$ and $z=6.5$).
\begin{figure*}
\includegraphics[angle=-90,width=15cm]{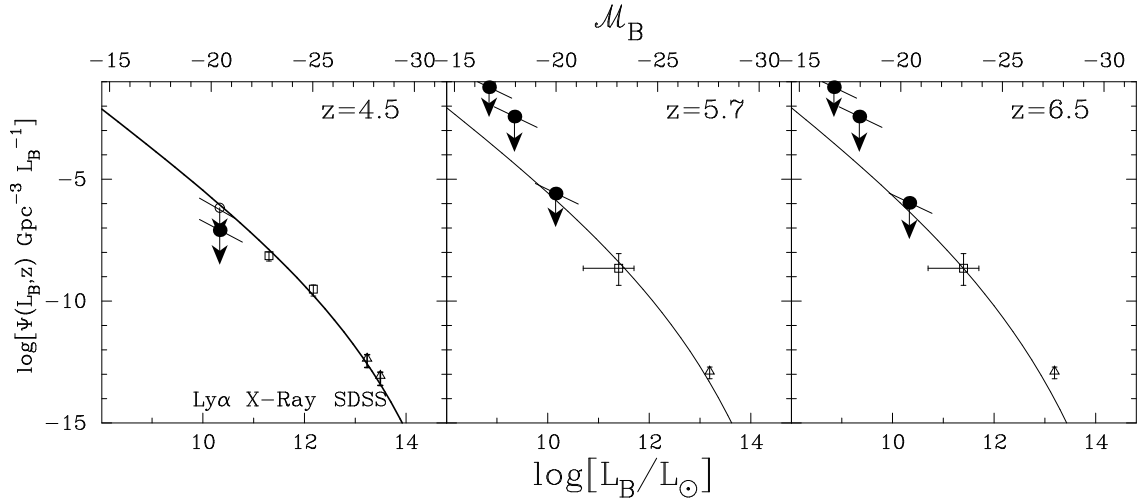}
  \caption{Ly$\alpha$ constraints on the quasar B-band luminosity function (QBLF), $\Psi(L_B,z)$ at redshifts $z=4.5$ ({\it left panel}), $z=5.7$ ({\it middle panel}) and $6.5$ ({\it right panel}), plotted as {\it filled circles} at log$[L_B/L_{\rm B,\odot}]\lsim 10$. The error-bars on the Ly$\alpha$ constraints denote $95\%$ confidence limits. The {\it open circle} at $z=4.5$ denotes the upper limit obtained from spectroscopic data alone (\S~\ref{sec:spec}).
  The SDSS and X-Ray data are taken from Fan et al (2001,2001b) ({\it open triangles}) and \citet{Barger03} ({\it open squares}), respectively. This figure shows that existing Ly$\alpha$ surveys allow the QBLF to be constrained at B-band luminosities as low as log$[L_B/L_{\rm B,\odot}]=8.5$ (or at absolute magnitude $M_B\sim -16$). Our constraint at $z=4.5$ suggests that the QBLF flattens for log$[L_B/L_{B,\odot}] \lsim 11$ (also see Fig.~\ref{fig:dPdB}).}
\label{fig:lumfunc}
\end{figure*} 
In Figure~\ref{fig:lumfunc} we compare the above constraint on the
quasar B-band luminosity function (denoted by the {\it filled circles} at
log$[L_B/L_{\rm B,\odot}]\lsim 10$ in central and right  panel),
 with data at higher
luminosities. Each panel corresponds to the redshift printed in the
upper right corner of the figure. The {\it open triangles} at
log$[L_B/L_{\rm B,\odot}]>13$ represent data from Fan et al
(2001b). The {\it open squares} are derived from
the X-Ray data presented by \citet{Barger03}, who plot 
the number density of AGN as a function of 
redshift in the rest frame soft X-Ray (E=0.5-2.0 keV) luminosity range
$L_X=10^{43}-10^{44}$ ergs s$^{-1}$ (hereafter, the 'faint bin') and
$L_X=10^{44}-10^{45}$ ergs s$^{-1}$ (hereafter, the 'bright bin'). For
the \citet{Sazonov04} template spectrum, the ratio of the rest frame
soft X-Ray (denoted by $L_X$) to blue band luminosity is $L_X/L_B\sim
0.5$. The faint and bright bins of \citet{Barger03} therefore constitute the
range $L_B=5\times 10^{10}-5\times 10^{11}L_{B,\odot}$ and $L_B=5\times 10^{11}-5\times 10^{12}L_{B,\odot}$, respectively. To convert the X-Ray number density to the luminosity function, $\Psi(L_B,z)$, we simply divided $n_{\rm faint}$ by the total luminosity width of the bin. We use $n_{\rm faint}=(1.0^{+3.0}_{-0.8}) \times 10^{-6}$ Mpc$^{-3}$ in both bins.

The error-bars on our constraints reflect the uncertainty in the exact ratio of $L_B/L_{{\rm Ly}\alpha}$ (Eq.~\ref{eq:key}) and denote the $95\%$ confidence
levels. Note that upper limits on $\Psi(L_{\rm B,min},z)$ as a function
of $L_{\rm B,min}$ have power law slopes of $-1$ (see
Eq.~\ref{eq:psinum}). The constraints from the wide field surveys lie on or slightly below the model predictions of \citet{Wyithe03} (shown as the {\it solid lines}). This implies that the absence of $z=5.7$ and $z=6.5$ AGN in wide-field narrow-band surveys rules out this model at $\geq 95\%$ confidence levels at both redshifts, at these low luminosities.

\subsubsection{Constraints on $\Psi(L_B,z)$ at $z=4.5$.}
\label{sec:selcon}
The above constraints at $z>5$ assumed that $\beta_l=1.66$. 
However, as will be discussed below, at $z=4.5$, extrapolation of this powerlaw from the X-Ray constraint at log$[L_{B}/L_{B,\odot}\sim 11]$ to $L_{\rm B,min}$, would have resulted in a number density of AGN that lies above the implied upper limit. To better constrain the luminosity function, $\Psi(L_{\rm B.min},z)$, and its slope at $z=4.5$, we combine the upper limit on the AGN number density from the Ly$\alpha$ survey with the AGN number densities derived from X-ray data \citep{Barger03,Cowie03}. We use $n_{\rm bright}=(1.0 \pm 0.5) \times 10^{-6}$ Mpc$^{-3}$ and $n_{\rm faint}=(3.3 \pm 1.3) \times 10^{-6}$ Mpc$^{-3}$, which we obtained by interpolating between the $z=3.5$ and $z=5.7$ data points of \citet{Barger03}.

We assume the luminosity function, $\Psi(L_B,z)$ to be 
of the form $k L^{-\beta_l}$. The number density of
AGN in the faint X-ray bin, $n_{\rm faint}$,
 fixes $k$ as a function of $\beta_l$.
Extrapolating this luminosity function to lower luminosities,
gives us the expected number density of AGN in the Ly$\alpha$ bin,
$n_{\rm AGN}=$ $k\int_{L_{\rm B,min}}^{L_{\rm B, max}}dL\hs L^{-\beta_l}$.
One of the selection criteria \citet{Dawson04} use to select $z=4.5$ Ly$\alpha$ emitter candidates states that a $z=4.5$ candidate must not be detected in the $B_W$ band of the NOAO deep wide-field survey \citep{Malhotra02}. We show in \S~\ref{sec:broadagn} that this implies that $L_{\rm B,max}\sim 2 \times 10^{11}L_{B,\odot}$. For a fixed $\beta_l$, we obtain a unique $n_{\rm AGN}$. Since $n_{\rm AGN}$ is only known as an upper limit, this allows us to put an upper limit on $\beta_l$ as follows.

\begin{figure*}
\includegraphics[angle=-90,width=15cm]{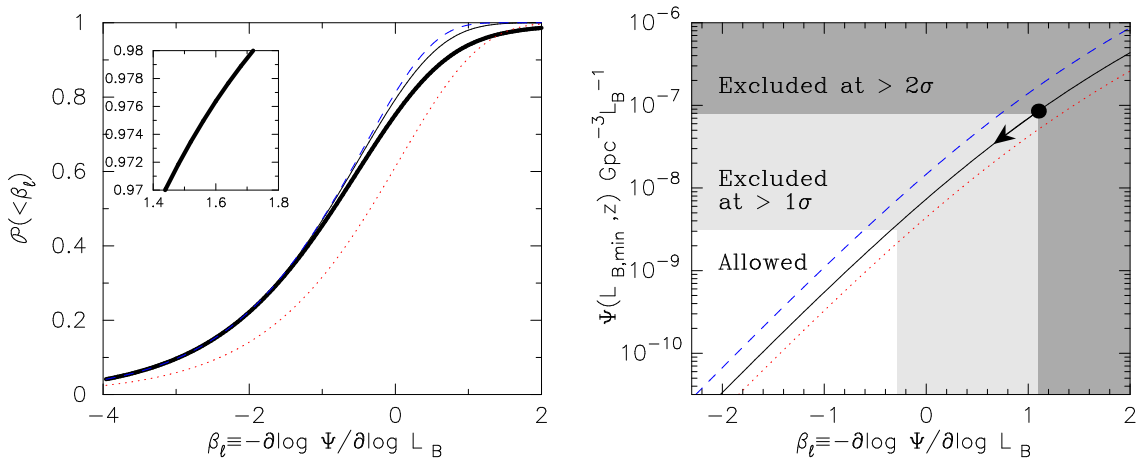}
  \caption{{\it Left Panel:} The probability $P(<\beta_l)$ that
  $-$\dPsidL \hs is less than $\beta_l$. The {\it thin black solid
  line} represents the constraint for the fiducial model with $L_{\rm
  B,min}=L_{{\rm Ly}\alpha{\rm,c}}$ and $n_{\rm faint}=3.3 \times
  10^{-6}$ Mpc$^{-3}$, for which $P(< 0.72)=0.95$. The effect of lowering
  $n_{\rm faint}$ ({\it red dotted line}, $n_{\rm faint}=2.0 \times
  10^{-6}$ Mpc$^{-3}$) and $L_{\rm B,min}$ ({\it blue
  dashed line}, $L_{\rm B,min}$=$0.50L_{{\rm Ly}\alpha{\rm,c}}$) is to
  reduce and increase $P(< \beta_l)$, respectively. The {\it thick solid
  line} shows the probability, $\mathcal{P}_{\rm marg}(<\beta_l)$, after marginalizing over $L_{\rm B,min}$ and $n_{\rm faint}$. We find marginal evidence ($75\%$ confidence level) that the luminosity function falls towards lower luminosities ($\mathcal{P}_{\rm marg}(< 0.0)=0.75$) below log$[L_B/L_{B,\odot}]=11$.
{\it Right Panel:} The expected value of
  $\Psi(L_{\rm B,min},z)$ is plotted as a function of $\beta_l$ for the models shown in the {\it left panel}.  
  The marginalised $2-\sigma$ upper limit (also shown in Fig~\ref{fig:lumfunc}) is shown as the {\it large filled circle}, which implies log$[\Psi(L_{\rm B,min},z)]$ $<-7.1$. The light and dark grey areas demarquate the region of $(\beta_l,\Psi)$-space excluded at the $>$ 68\% (1-$\sigma$) and $> 95\%$ (2-$\sigma$) confidence level, after marginalising over $n_{\rm faint}$ and $L_{\rm B,min}$. For reference, the model of \citet{Wyithe03} predicts log$\Psi \sim -5.7$.}
\label{fig:dPdB}
\end{figure*}

In the absence of AGN within the volume probed by the Ly$\alpha$
survey, the probability that the true number lies between $\mu$ and $\mu+d\mu$ is given by $dP=$exp$(-\mu)d\mu$ (where we have assumed a Poisson distribution). This may be recast as

\begin{equation}
P(<\mu)=1-{\rm e}^{-\mu},
\end{equation} 
which gives the probability that the expected number of AGN within the
survey volume is less than $\mu$. The expected number of AGN is
given by

\begin{equation}
\mu=V_{\rm survey }\int_{L_{\rm B,min}}^{L_{\rm B,max}} dL \hs\Psi(L,z).
\end{equation}  For a fixed $L_{\rm B,max}$, we find that $\mu$ is a function of $L_{\rm B,min}$, $n_{\rm faint}$ and $\beta_l$ only. In the {\it left panel} of Figure~\ref{fig:dPdB} the {\it thin black solid line} shows the probability $P(< \beta_l)$ as a function of $\beta_l$, for the fiducial values of $L_{\rm B,min}=L_{{\rm Ly}\alpha{\rm,c}}$ and $n_{\rm faint}=3.3 \times
10^{-6}$ Mpc$^{-3}$. In this case $P(< 0.72)=0.95$, i.e. the
slope is $\beta_l \lsim 0.72$ at the $95 \%$ confidence level. To
illustrate the dependence of $P(< \beta_l)$ on $L_{\rm B,min}$, and
$n_{\rm faint}$, we have also plotted $P(< \beta_l)$ for $(n_{\rm
faint}, L_{\rm B,min})=(2.0 \times 10^{-6}$ Mpc$^{-3}$, $L_{{\rm
Ly}\alpha{\rm,c}})$ ({\it red dotted line}) and $(3.3 \times 10^{-6}$
Mpc$^{-3}$, $0.5 L_{{\rm Ly}\alpha{\rm,c}})$ ({\it blue dashed
line}). The $2-\sigma$ upper limit on $\beta_l$ increases with
decreasing $n_{\rm faint}$ and increasing $L_{\rm B,min}$. To
eliminate the dependence of $P(< \beta_l)$ on $L_{\rm B,min}$ and
$n_{\rm faint}$, we marginalise over these parameters

\begin{eqnarray}
\mathcal{P}_{\rm marg}(< \beta_l)=\int dn_{\rm faint}\int dL_{\rm B,min}\frac{dP}{dn_{\rm faint}}\frac{dP}{dL_{\rm B,min}} \times \nonumber \\
\times [1-{\rm e}^{-\mu(n_{\rm faint},L_{\rm B,min},\beta_l})],
\end{eqnarray} 
where $dP/dn_{\rm faint}$ and $dP/dL_{\rm B,min}$ are the probability
distributions for $n_{\rm faint}$ and $L_{\rm B,min}$,
respectively. We choose $dP/dn_{\rm faint}$ to be Gaussian with
$n_{\rm faint}=(3.3 \pm 1.3)\times 10^{-6}$ Mpc$^{-3}$, and $dP/dL_{\rm
B,min}$ to be log normal with log$(L_{\rm B,min}/L_{B,\odot})=10.3 \pm
0.2$ (this range is motivated by the distribution in the ratio
$L_{{\rm Ly}\alpha}/L_B$, found from Eq.~\ref{eq:key} in
\S~\ref{sec:lum}). The marginalised constraint is given by the {\it
thick solid line} in Figure~\ref{fig:dPdB}, which shows that $\beta_l
\lsim 1.1$ at the $95 \%$ confidence level. The {\it right panel} of
Figure~\ref{fig:dPdB} shows the expected value of $\Psi(L_{\rm
B,min},z)$ as a function of $\beta_l$. The constraint $\beta_l < 1.1$
yields log$[\Psi(L_{\rm B,min},z)]$ $<-7.1$ and is shown
as the {\it filled circle}.
This $2-\sigma$ upper limit on $\Psi(L_{\rm B,min},z)$ at $z=4.5$ is also shown in the left panel of Figure~\ref{fig:lumfunc} as the {\it filled circle}. The {\it open triangles} at log$[L_B/L_{\rm B,\odot}]>13$ represent data from Fan et al (2001) and the {\it open squares} represent the X-ray constraints derived from the data presented by Barger et al. (2003) and Cowie et al. (2003). Figure~\ref{fig:lumfunc} implies that at $z=4.5$, the QLF flattens below log$(L_B/L_{B,\odot})\sim 11$, as may also be seen via the model quasar luminosity function from Wyithe \& Loeb 2002 and 2003 (plotted as the {\it black solid line} in Fig.~\ref{fig:lumfunc}). 
It is worth pointing out how the exact marginalised constraint on $\beta_l$ depends on the assumed survey sensitivity. If we decrease/increase $f_{\rm lim}$ by a factor of 2 then $\beta_l < 0.8/1.4$ at the $95\%$ confidence level, respectively. The value of $\mathcal{P}_{\rm marg}(< 0.0)$ barely depends on $f_{\rm min}$. The decreased sensitivity corresponds approximately to the sensitivity limit reported by \citet{Dawson04}, while the increased sensitivity corresponds to the sensitivity derived from their actual faintest Ly$\alpha$ detections. 

The constraints obtained so far only used the non-detection of broad line emitters in the narrow band survey. \citet{Dawson04} also obtained spectra of 25 of their candidate LAEs. Below, we constrain $\Psi(L_{\rm B},z)$ using this spectroscopic data.
\subsection{Constraints on $\Psi(L_B,z)$ at $z=4.5$ from Spectroscopic Follow-Up.}
\label{sec:spec}
\citet{Dawson04} obtained follow-up high resolution spectra for 25 candidate LAEs. Of these 25 candidates, 18 were genuine $z=4.5$ objects and the other 7 were either not detected (6) or a lower redshift [O II] emitter (1). All 18 $z=4.5$ objects were identified as galaxies, based on 1) the observed narrow physical line widths of the Ly$\alpha$ lines ($\Delta v \lsim 500$ km s$^{-1}$, also see \S~\ref{sec:lyaagn}), and 2) the lack high--ionization state UV emission lines, symptomatic of AGN activity, in their spectra. If the sample of 350 detections contains less than 10\% AGN, $f_{\rm AGN}\lsim 0.1$, then the probability that a random sample of 25 contains no AGN is $\sim(1-f_{\rm AGN})^{25}\gsim 0.05$ (In \S~\ref{sec:Xray} we show that $f_{\rm AGN}\lsim 0.1$ is consistent with the constraints imposed by the X-Ray observations of Wang et al. 2004). From the spectroscopic data alone, we can therefore put a $2-\sigma$ upper limit on the number of AGN within the original survey volume ($1.5 \times 10^6$ Mpc$^3$) at $N_{\rm AGN} \lsim 350f_{\rm AGN}=35$ down to a Ly$\alpha$ flux of $1.1 \times 10^{-17}$ ergs s$^{-1}$ cm$^{-2}$. Using Eq~(\ref{eq:psinum}) we obtain an upper limit on $\Psi(L_B,z)$, which is shown as the {\it open circle} in Figure~\ref{fig:lumfunc}. This constraint on $\Psi(L_{\rm B},z)$ is weaker than our original constraint, but still rules out the model luminosity function at $> 95\%$. 

\section{The Nature of Very Faint AGN and Their Observable Properties.}
\label{sec:phys}

We have demonstrated how Ly$\alpha$ surveys may be used to constrain
the (very) faint end of the quasar luminosity function. This
constraint is purely empirical and we stress that so far no details of
the process of gas accretion onto black holes have been assumed. The
following discussion focuses on the nature of faint AGN (with
log$[L_B/L_{\rm B,\odot}]\lsim 10$). First, we present a more general
calculation of the observable Ly$\alpha$ properties of AGN. The result
of this calculation allows us to determine the range of black hole
masses probed by Ly$\alpha$ surveys and may be used to identify AGN
among known Ly$\alpha$ emitters, especially in combination with our
estimates for the observable properties of these faint AGN in the
optical (\S~\ref{sec:broadagn}) and X-ray (\S~\ref{sec:Xray}) bands.

\subsection{Observable Ly$\alpha$ Properties of Faint AGN}
\label{sec:lyaagn}

First we discuss the intrinsic Ly$\alpha$ properties of AGN. The Kaspi relation \citep{Kaspi00,Peterson04} relates the mass of the black hole powering the quasar, $M_{\rm BH}$, to its continuum luminosity at 5100 \AA
\begin{equation}
M_{\rm BH}=7.6\pm 1.3 \times 10^7 \Big{(} \frac{\lambda F_{\lambda}
(5100\AA)}{10^{44} {\rm erg} \hs {\rm s}^{-1}}\Big{)}^{0.79}\hs
M_{\odot}.
\label{eq:kaspi}
\end{equation}
We define the ratio $\mathcal{R} \equiv [\lambda F_{\lambda}$(1200
\AA)]/$[\lambda F_{\lambda}$(5100 \AA)]= $[E F_{E}$(1200 \AA)]/$[E
F_{E}$(5100 \AA)] (see \S~\ref{sec:lum}). The template given by
\citet{Sazonov04} yields $\mathcal{R}=1.8$. For a given Ly$\alpha$ EW,
the Ly$\alpha$ luminosity is therefore uniquely determined by $M_{\rm
BH}$ through
\begin{equation}
L_{{\rm Ly}\alpha,43}=1.6 \hs\Big{(}\frac{{\rm EW}}{100 {\rm \AA}} \Big{)}\Big{(}\frac{M_{\rm BH}}{7.6 \times 10^7 M_{\odot}}\Big{)}^{1.3}, 
\label{eq:lyalum}
\end{equation} 
where $L_{{\rm Ly}\alpha,43}$ is in units of $10^{43}$ ergs
s$^{-1}$. The Ly$\alpha$ emission lines from AGN are broader than
those of galaxies. \citet{Kaspi00} and \citet{Peterson04} found the Half Width at Half Maximum (HWHM) of the Balmer lines to be related to the continuum 
luminosity at 5100 \AA. If the same relation holds for the Ly$\alpha$ line, then its HWHM is given by
\begin{equation}
v_{\rm HWHM}=2 \times 10^{3}\Big{(}\frac{M_{\rm BH}}{7.6 \times 10^7 M_{\odot}}\Big{)}^{-0.34} \hs {\rm km}\hs {\rm s}^{-1}.
\label{eq:vfwhm}
\end{equation} 
Note that the observations presented in \citet{Kaspi00} (their Figure
7) suggest that $v_{\rm HWHM}$ does not increase beyond $v_{\rm
HFHM}\sim 3 \times 10^3$ km s$^{-1}$ at low masses.

To obtain the AGNs' observed properties in the Ly$\alpha$ line from
the above relations, we require knowledge of the fraction of transmitted
Ly$\alpha$-flux through the IGM as a function of frequency. We first
focus on the observed Ly$\alpha$ properties of AGN at $z=4.5$ and
$z=5.7$. Although the universe is fully reionised at $z < 6$
\citep[e.g.][]{Fan02}, a trace quantity of neutral hydrogen is sufficient
to scatter Ly$\alpha$ blueward of the Ly$\alpha$ line center out of
our line of sight. For this reason the IGM, to first order, erases the
blue half of the Ly$\alpha$ line, so that the IGM transmission is
$\mathcal{T}=0.5$. Various refinements of this scenario are
possible. Infall of the IGM around massive objects causes the IGM to
erase a part of the red side of the Ly$\alpha$ line as well
\citep{Barkana04}, which reduces $\mathcal{T}$. On the other hand, the
proximity effect around these fainter AGN increases
$\mathcal{T}$. These two effects counteract each other, and ignoring
both does not add a significant error to our estimate of
$\mathcal{T}$. This is especially true when considering the large width of
the Ly$\alpha$ line emitted by AGN. When the IGM erases the
 blue half of the Ly$\alpha$ line, the observed
$v_{\rm HWHM}$ is reduced by a factor of 2 relative to the value in
Eq~(\ref{eq:vfwhm}). Note that $v_{\rm HWHM}=1500$ km s$^{-1}$ corresponds
to an observed FWHM of $(1+z)2v_{\rm HWHM}/c \sim 66\AA$ at $z=4.5$
 (this value was used in \S~\ref{sec:numden}). 
Note that one has to be careful not to confuse the observed FWHM 
of the Ly$\alpha$ line with its EW (it may be particularly confusing
since the observed rest-frame EW of AGN are $50$ \AA, comparable 
to the value of the observed FWHM here).

At $z=6.5$, the IGM is believed to contain a significant fraction of neutral
hydrogen \citep{Wyithe04,Mesinger04,Fan06}. For Ly$\alpha$
sources embedded in a neutral IGM, the damping wing of the
Gunn-Peterson trough can extend to the red side of the line and erase
a significant fraction of the total Ly$\alpha$ flux. However, 
this effect is reduced when the Ly$\alpha$ source is
surrounded by an HII region \citep{CenHaiman00,MadauRees00}. Moreover,
 sources with sufficiently broad emission lines ($v_{\rm HWHM}>300$ km
s$^{-1}$) can remain detectable even in the absence of such
an HII region \citep{Haiman02}. In Appendix~\ref{app:trans} we show
that a representative number for the IGM transmission $\mathcal{T}$
for faint AGN embedded in a neutral IGM at $z=6.5$ 
is $\mathcal{T}=0.3$ and that the observed $v_{\rm HWHM}$
is reduced, again by a factor of $\sim$2 relative to the value in
Eq~(\ref{eq:vfwhm}). Furthermore, we show that the Gunn Peterson damping wing may cause the observed line center of AGN at $z=6.5$ to be redshifted by up to $\sim 2000$ km s$^{-1}$ relative to other emission lines (see Appendix~\ref{app:trans}).

Next, we write the total detectable Ly$\alpha$ flux of AGN as
\begin{eqnarray}
f_{\alpha}&=&\frac{L_{\alpha}}{4 \pi d_L^2(z)}\mathcal{T}\nonumber\\
&\approx& 4 \times 10^{-17}\Big{(}\frac{M_{\rm BH}}{7.6 \times 10^7} \Big{)}^{1.3}
\nonumber \\ 
&\times& \Big{(}\frac{\mathcal{T}}{0.5}\Big{)}\Big{(}\frac{{\rm EW}}{100 \AA} \Big{)} \Big{(}\frac{1+z}{6} \Big{)}^{-2.8} \hs {\rm erg} \hs {\rm s}^{-1} \hs {\rm cm}^{-2}
\label{eq:fa}
\end{eqnarray} 
in which $\mathcal{T}\sim 0.5$ for $z < 6$ and $\mathcal{T}\sim 0.3$
for $z=6.5$. To allow $f_{\alpha}$ to be written as a simple function 
of $(1+z)$, we approximated the luminosity
distance by $d_L(z)=$ $3.6 \times 10^{4}([1+z]/5)^{1.4}$ Mpc, which
is within $< 6\%$ of its actual value in the range $z=3-10$.
Eq~(\ref{eq:fa}) shows that existing Ly$\alpha$ surveys
could have detected AGN powered by black holes with masses of $M_{\rm
BH}\sim 10^6-10^7 M_{\odot}$ at $z=4.5-6.5$. 

Equation~(\ref{eq:fa}) assumed that the Kaspi-relations derived from
observations of luminous AGN also relate the black-hole
mass to the Ly$\alpha$ luminosity for low luminosity
AGN. Alternatively, if we assume that faint AGN are powered by less
massive black holes accreting at their Eddington limit, then these
black-holes are up to $\sim 5$ orders of magnitude less massive than those
of the observed SDSS quasars, yielding $M_{\rm BH}\sim$ a few $10^4
M_{\odot}$. Combined with the above argument, we therefore conclude
that existing Ly$\alpha$ surveys could have detected AGN powered by
black holes with masses in the range $M_{\rm BH}=10^4-10^7 M_{\odot}$
at $z=4.5-6.5$. Here, the lower end of this mass range is probed only by
the spectroscopic surveys of gravitationally lensed regions.
\begin{figure*}
\includegraphics[width=8cm,angle=270]{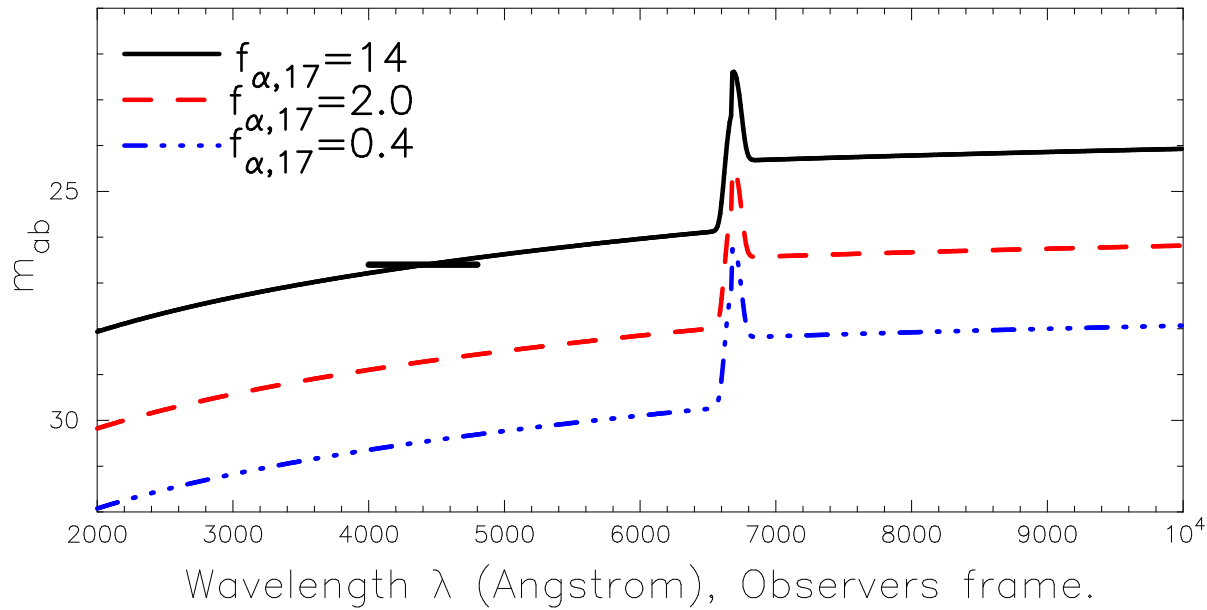}
 \caption{Model spectra of faint AGN at $z=4.5$ with observable
 Ly$\alpha$ flux, $f_{\alpha}=1.4 \times 10^{-16}$ ({\it black solid
 line}), $2 \times 10^{-17}$ ({\it red dashed line}) and $4\times
 10^{-18}$ ({\it blue dotted line}) ergs s$^{-1}$ cm$^{-2}$ obtained with Eq.(~\ref{eq:broadagn}) with EW=100 \AA. This figure shows
 that AGN with $f_{\alpha} \gsim 2 \times 10^{-16}$ ergs s$^{-1}$
 cm$^{-2}$ have a $B_W$ apparent magnitude of $m_{AB}\lsim
 26.6$. These AGN would have been detected in the NOAO deep field
 (whose detection limit is indicated by the {\it thick horizontal
 line}) and would therefore have been discarded as
 being candidate high redshift Ly$\alpha$ emitters
 \citep{Malhotra02}. Furthermore, the AGN with
 $f_{\alpha}= 4 \times 10^{-18}$ ergs s$^{-1}$ cm$^{-2}$, which
 corresponds to \citet{Taniguchi05}'s detection limit, have 
 apparent magnitudes $m_{AB}\gsim 28$, which makes them difficult
to detect at other wavelengths.}
\label{fig:mab}
\end{figure*}
\subsection{AB-magnitudes of Faint AGN}
\label{sec:broadagn}

Using the template spectrum for AGN given by \citet{Sazonov04}, we
calculate the apparent AB-magnitude of faint AGN as a function of
observed wavelength. The AB-magnitude of an object with flux density
$F_{\nu}$ (in ergs s$^{-1}$ cm$^{-2}$ Hz$^{-1}$) is
\begin{equation}
m_{AB}=-48.6-2.5\log(F_{\nu}).
\end{equation} 
At rest frame energies $E < 10$ eV and $E > 10$ eV, the AGN continuum
follows the power-laws $F_{\nu} \propto \nu^{-0.6}$ and $F_{\nu}
\propto \nu^{-1.7}$, respectively \citep{Sazonov04}.
 For an AGN whose Ly$\alpha$ flux is
$f_{\alpha,17}\times 10^{-17}$ ergs s$^{-1}$ cm$^{-2}$, the apparent
magnitude may then be written as
\begin{equation}
m_{AB}(\lambda)=26.4-2.5\Big{[}\log\Big{(}\frac{5.5}{1+z} \Big{)}+\log
\mathcal{K}(\lambda)\Big{]}, \nonumber
\end{equation}
where
\begin{eqnarray}
{\mathcal K}(\lambda)&=&
\Big{(}\frac{f_{\alpha,17}}{2} \Big{)}
\Big{(}\frac{100 \AA}{{\rm EW}} \Big{)}
\Big{(}\frac{0.5}{\mathcal{T}} \Big{)}
\Big{(}\frac{\lambda}{\lambda_{\alpha,z}}\Big{)}^{0.6} \hs \lambda > \lambda_{\alpha,z}, \nonumber\\
&=&\langle {\rm e}^{-\tau} \rangle
\Big{(}\frac{f_{\alpha,17}}{2} \Big{)}
\Big{(}\frac{100 \AA}{{\rm EW}} \Big{)}
\Big{(}\frac{0.5}{\mathcal{T}} \Big{)}
\Big{(}\frac{\lambda}{\lambda_{\alpha,z}}\Big{)}^{1.7} 
\hs \lambda < \lambda_{\alpha,z}.\nonumber\\
\label{eq:broadagn}
\end{eqnarray} 
Here $\lambda_{\alpha,z}$ is the redshifted Ly$\alpha$ wavelength
($\lambda_{\alpha,z}=1216[1+z]$ \AA). 

In Figure~\ref{fig:mab} we plot
$m_{AB}$ as a function of observed wavelength, $\lambda$ in the range
$\lambda=2000-10000$ \AA \hs for three values of the total flux in the
Ly$\alpha$ line, $f_{\alpha,17}=14$ ({\it black solid line}), $2$
({\it red dashed line}) and $0.4$ ({\it blue dotted line}), all at
$z=4.5$. We assumed the Ly$\alpha$ equivalent width to be EW$=140$
\AA. The break at $\lambda=6700$ \AA \hs is the Lyman break caused by
the IGM; the Ly$\alpha$ forest reduces the mean flux blueward of the
Ly$\alpha$ line by a factor of $\langle {\rm e}^{-\tau} \rangle$. This
factor has been determined observationally and is $\langle {\rm
e}^{-\tau} \rangle=0.25$ at $z=4.5$
\citep[e.g.][]{Fan02}. Figure~\ref{fig:mab} (and Eq.~18) shows that
AGN with $f_{\alpha} \gsim f_{\rm max}= 1.4 \times 10^{-16}$ ergs
s$^{-1}$ cm$^{-2}$ have a $B_W$ apparent magnitude of $m_{AB}\lsim
26.6$ ($\lambda \sim 4400$ \AA). These AGN would have been detected at the $\gsim 5-\sigma$ level in the $B_W$ band of the NOAO deep wide-field survey \citep[][this detection limit is indicated by the {\it thick horizontal
 line} at $m_{AB}=26.6$]{Januzzi99}. Since one of the selection criteria \citet{Dawson04} use to select $z=4.5$ Ly$\alpha$ emitter candidates states that a $z=4.5$ candidate must not be detected in the $B_W$ band of the NOAO deep wide-field survey \citep{Malhotra02}, AGN with $f_{\alpha} \gsim  f_{\rm max}=1.4 \times 10^{-16}$ ergs s$^{-1}$ cm$^{-2}$ would not have made it into the sample. Therefore, with the current selection criteria, the Ly$\alpha$ survey performed by \citet{Dawson04}
only probes AGN in the luminosity range $L_{\rm B} \in [L_{\rm B,min},L_{\rm B,max}]$, with $L_{\rm B,max}\sim 2 \times 10^{11}L_{B,\odot}$ (this luminosity was used in \S~\ref{sec:selcon}).

\subsection{X-Ray Emission from Faint AGN}
\label{sec:Xray}

X-Rays provide a reliable pointer to AGN activity.  Using the
\citet{Sazonov04} template, we calculate the ratio of detectable
Ly$\alpha$ and X-Ray flux (in the observed 0.5-10.0 keV band)
\begin{equation}
\frac{f_{\alpha}}{f_X}\sim 0.05 \Big{(}\frac{\mathcal{T}}{0.5}\Big{)}\Big{(}\frac{{\rm EW}}{100\hs \AA}\Big{)}.
\end{equation} 
This ratio changes by less than $\sim 10\%$ between $z=4.5-6.5$. The
total X-Ray flux in the observers' 0.5-10.0 keV band is $\sim 20$
times higher than the total Ly$\alpha$ flux. For example, an AGN with
a detected Ly$\alpha$ flux of $2 \times 10^{-17}$ ergs s$^{-1}$ cm$^{-2}$ is
expected to have an X-ray flux of $\sim 4 \times 10^{-16}$ ergs s$^{-1}$
cm$^{-2}$. This lies slightly below the quoted 3$-\sigma$
detection limit in the X-ray observations of individual Ly$\alpha$
emitter candidates presented by \citet{Wang04} ($\sim 6.6 \times 10^{-16}$ ergs s$^{-1}$ cm$^{-2}$ in the 0.5-10.0 keV band). Therefore, even if AGN were in 
the sample, these X-ray observations would not necessarily have revealed them (\S~\ref{sec:numden}). Also, provided the fraction of AGN is $f_{\rm AGN} \lsim 0.1$ (\S~\ref{sec:spec}), these would not have appeared in the stacked X-ray image, since stacking 101 images increases the noise by $101^{1/2}$, whereas the X-Ray signal from AGN would increase by $f_{\rm AGN}\times 100\lsim 10$. The signal--to--noise ratio would thus be conserved at best.

Similarly, the total X-Ray flux in the observers' 0.5-2.0 keV band is
$\sim 4$ times higher than the total Ly$\alpha$ flux. It may come as a
surprise then, that our constraints are fainter than those inferred
from the {\it Chandra Deep Field}. The X-Ray detection threshold in
the {\it Chandra Deep Field North} in the 0.5-2.0 keV band is $\sim
1.5 \times 10^{-17}$ ergs s$^{-1}$ cm$^{-2}$, which would correspond
to a Ly$\alpha$ flux of $\sim 3\times 10^{-18}$ ergs s$^{-1}$
cm$^{-2}$, well below the sensitivity limit of $f_{\rm lim}\sim 1.1
\times 10^{-17}$ ergs s$^{-1}$ cm$^{-2}$ in the survey performed by
\citet{Dawson04}. However, estimates of the photometric redshifts 
for the X-Ray sources require that they be bright
enough to be detected in the Subaru Sloan z' band (which is centered
on $\lambda\sim 9000$ \AA). 
This imposes a flux limit corresponding to
AB magnitudes of the X-Ray sources in the Sloan z' band that are below
$\lsim 25.2$. Using the quasar template, we find that this excludes AGN with
B-Band luminosities less than $6 \times 10^{10} L_{B,\odot}$, which
agrees very well with the lower luminosity bound of the 'faint' X-ray
bin in \S~\ref{sec:selcon}.

The previous discussion demonstrates an observational bias
against identification of high redshift faint AGN in deep X-Ray
observations. Within narrow redshift windows, this bias could be
alleviated by combining X-Ray observations with deep Ly$\alpha$
observations, since the detection of a Ly$\alpha$ line would determine
the AGN's redshift. Similarly, the bias against identifying bright AGN
in wide field Ly$\alpha$ surveys (\S~\ref{sec:broadagn}) could be
alleviated in combination with deep X-ray observations. These points
illustrate the utility of combining deep X-Ray and Ly$\alpha$
observations to identify faint AGN, provided the sensitivities in each
observation probe the same population of AGN.
\section{Discussion}
\label{sec:discuss}

We have shown that at $z=4.5$ and below $M_{\rm B}\sim-20$, the quasar
B-band luminosity function rises more slowly towards lower luminosities,
\dPsidL$\gsim -1.1$ (95\% confidence level), than has been observed
 at higher luminosities and lower redshifts, where \dPsidL\hs $\sim -1.6$
 \citep{Pei95,Boyle00}. This flattening of the faint end of the luminosity
function towards higher redshift is consistent with the recent work
by \citet{Hunt04}, who found that \dPsidL$=-1.24 \pm 0.07$ at $z=3$.
We find marginal evidence ($75\%$ confidence level) that,
 in fact, the luminosity function falls towards lower luminosities below $M_{\rm B}\sim-20$. This finding is in contrast to observations at higher luminosities. Moreover the observed number counts lie well below the model predictions of
Wyithe \& Loeb 2003 (Fig.~\ref{fig:lumfunc}).
 This may be explained in three ways: 

1) Our work has focused on the luminosity function of broad-lined AGN (Type I), which in the unified model for AGN are the same as Type II AGN (e.g. Norman et al, 2002), but unobscured by the thick absorbing torus. \citet{Simpson05} has shown that the fraction of type I AGN increases with luminosity, which is supported theoretically by the 'receding torus' model \citep[e.g.][]{Lawrence91}. This would imply that a luminosity function that incorporates both type I and II AGN does not flatten as much at $z=4.5$ as shown in Figure~\ref{fig:lumfunc}. However, to fully explain the observed flattening of the luminosity function requires the fraction of type 1 AGN to decrease by $\sim 1-2$ orders of magnitude between log$[L_B/L_{B,\odot}]=10$ and $\sim 11.5$ . Since this luminosity dependence of the type I AGN fraction is much stronger than has been observed, this is very unlikely.

2)  Gas accretion onto black holes in the mass-range
$M_{\rm BH}=10^6-10^7 M_{\odot}$ is suppressed.
 One origin of this suppression may
be negative AGN feedback. AGN with log$[L_B/L_{\rm B,\odot}]=10.5$ are
typically powered by black holes in the mass range $M_{\rm BH}\sim
10^{6} M_{\odot}$ to a few $10^{7} M_{\odot}$
(\S~\ref{sec:lyaagn}). According to the relation between $M_{\rm BH}$
and the circular velocity $v_{\rm circ}$ of the dark matter halo that
hosts the black hole [\citet{Ferrarese00,Ferrarese02}], this corresponds
to $v_{\rm circ}=100-250$ km s$^{-1}$. For comparison, suppression of
accretion due to a photoionised IGM at $z=4.5-6.5$ only occurs at
$v_{\rm circ}=40-60$ km s$^{-1}$ \citep[e.g.][]{Feedback}. However,
\citet{Dekel86} have shown that supernova driven gas loss as a result
of the first burst of star formation becomes significant in halos with
$v_{\rm circ} \lsim 100$ km s$^{-1}$. The latter feedback mechanism
would therefore provide a more plausible explanation for the reduced
gas accretion efficiency onto black holes in the mass range $M_{\rm
BH}\sim 10^{6} M_{\odot}-$ a few $10^{7} M_{\odot}$. 

3) The  number of black holes in the range $M_{\rm BH}=10^5-10^7
M_{\odot}$ is lower than expected from the $M_{\rm BH}-v_{\rm
circ}$-relation by $\sim$ two orders of magnitude. \citet{Haiman04} used the rareness of black holes with masses $M_{\rm BH} \lsim 10^7M_{\odot}$ as a possible explanation for their model of the luminosity function of radioloud quasars to overpredict the abundance of faint radio sources, by one-two orders of magnitude. The possible reduction in the number of black holes with masses $M_{\rm BH} \lsim 10^7M_{\odot}$, may reflect the existence of a minimum super massive black hole mass, as envisioned in some formation scenarios \citep[e.g.][in which the minimum black hole mass would be $M_{\rm BH}=10^6M_{\odot}$]{Haehnelt98}.

Our finding that at $z=4.5$ the QLF flattens more than
previously believed for log$[L_B/L_{B,\odot}]\lsim 11$ also implies that faint AGN contribute less photons to the ionising background than previously
thought. However, this is only a small effect since the total ionising
photon output from AGN per unit volume is $\propto \int
L\Psi(L,z)dL\propto \int L^{-\beta_l+1}dL$. For $\beta_l < 2$, this
integral is dominated by luminous AGN. A more intriguing implication
concerns miniquasars, which are quasars powered by black holes in the
mass range $M_{\rm BH}=10-10^5 M_{\odot}$. It has been suggested that
miniquasars could contribute significantly to the ionising background
at high redshift \citep{Madau04,Ricotti04}. However, if quasar
activity decreases with decreasing black hole mass (as our results
suggest), and if this trend continues into the miniquasar-realm, then
it follows that miniquasars would not be efficient producers of ionizing
radiation and would not have contributed significantly to the ionizing
background. It is worth emphasising that this miniquasar-realm may be accessible with existing deep spectroscopic surveys of gravitationally lensed regions \citep{Santos04lens}. In these surveys, black holes with masses of $M_{\rm BH}\sim 10^4M_{\rm BH}$ can be detected, provided these are accreting at their Eddington limit (\S~\ref{sec:lyaagn}). Currently, the only observational constraints on the abundance of high redshift miniquasars are derived from the cosmic X-Ray and infrared backgrounds (Dijkstra et al. 2004b, Salvaterra et al. 2005).
\section{Conclusions}
\label{sec:conclusion}

Recent Ly$\alpha$ surveys have detected Ly$\alpha$ emitting objects
from redshifts as high as $z=6.5$, and at luminosities as low as
$10^{41}$ erg s$^{-1}$ \citep[e.g.][]{Santos04lens}. 
No evidence of AGN activity exists among these several hundred Ly$\alpha$ emitters \citep{Dawson04,Wang04,Taniguchi05}. Wide field Ly$\alpha$ surveys are designed to deeply image wide fields on the sky, yielding survey volumes in a
narrow shell of redshift space as large as $10^5-10^6$ Mpc$^3$, while
deep spectroscopic surveys of gravitationally lensed regions probe deeper
into smaller volumes. The absence of AGN within these fields can place a tight upper limit on the number density of AGN with Ly$\alpha$ luminosities
exceeding the surveys detection thresholds, $L_{{\rm Ly}\alpha,{\rm
c}}$.

In \S~\ref{sec:lum} we have shown empirically that the Ly$\alpha$
luminosities of AGN equal their B-band luminosities to within a factor
of a few (Eq.~\ref{eq:key}). As a result, deep Ly$\alpha$ surveys can
be used to obtain upper limits on the number density of AGN with
B-Band luminosities exceeding $L_{\rm B,min}\sim 1.4 L_{{\rm
Ly}\alpha,{\rm c}}$. When expressed in B-band solar luminosities,
$L_{\rm B,min}$ is $\sim10^{8.5}L_{\rm B,\odot}$ (which
corresponds to an absolute magnitude of $M_B=-16$). In \S~\ref{sec:lyaagn} we demonstrated that such AGN are expected to be powered by black holes with masses in the range $M_{\rm BH}=10^4-10^7 M_{\odot}$ at $z=4.5-6.5$.

We derive upper limits on AGN number densities and constrain the quasar B-band
luminosity function $\Psi(L_{\rm B,min},z)$ at a luminosity $L_{\rm
B,min}$.  The non-detection of AGN among $z=4.5-6.5$ LAEs rules out the model predictions by Wyithe \& Loeb (2003), which succesfully reproduce the brighter end of the observed quasar luminosity function at $z=2-6$, at $\geq 95\%$ confidence levels at all redshifts. At $z=4.5$, we find that $\partial$log$\Psi/\partial$log$L_B \geq -1.6$, the value observed at lower redshifts, for log$[L_B/L_{B,\odot}] \lsim 11$ at the $98\%$ confidence level (Fig~\ref{fig:dPdB}). We find marginal evidence that at these luminosities, the luminosity function rises with luminosity, corresponding to a powerlaw slope $>0$ (75\% confidence level). In other words, the QLF may increase with $L_B$ at these faint luminosities,
in contrast to observations of more luminous AGN. These results
represent the faintest observational constraints on the quasar 
luminosity function at these redshifts to date.

We have found that models of the quasar luminosity function which are
successful in reproducing the bright end of the quasar luminosity
function predict more AGN to be present than are observed, by up to two
orders of magnitude (Fig.~\ref{fig:lumfunc}) at $z\sim4.5$. These results imply either that accretion onto lower mass black holes is less efficient than onto their more massive counterparts, or that the number of black holes powering AGN with $M_B\gsim-20$ is lower than expected from the $M_{\rm BH}-\sigma$ relation by one-two orders of magnitude. Extrapolating from reverberation-mapping studies suggests that these black holes would have $M_{\rm BH}=10^6-10^7 M_{\odot}$.

Our work has demonstrated the effectiveness of Ly$\alpha$ surveys in
constraining the faint end of the quasar B-band luminosity
function. Deeper and larger surveys will allow for a better determination
of its slope, and whether indeed the quasar luminosity function rises with luminosity for $M_B\gsim-20$ at $z=4.5$, and at other redshifts. These constraints will offer new insights on the growth of low mass black holes and their relation to the known super massive black holes. To help identify AGN among observed Ly$\alpha$ emitters, we have modeled the observable properties of the Ly$\alpha$ line for high redshift, faint AGN. Using the empirical
Kaspi-relations, we estimate that the observable Ly$\alpha$ line
widths (Half Width at Half Maximum) of faint AGN will be $\sim 1500$
km s$^{-1}$. For AGN embedded in a neutral medium the peak of the
Ly$\alpha$ line is redshifted up to $2000$ km s$^{-1}$ relative to the
true line center (\S~\ref{sec:lyaagn}). To facilitate the
identification of these faint AGN, we have estimated their observable
properties in the visible (\S~\ref{sec:broadagn}, Fig~\ref{fig:mab})
and X-Ray bands (\S~\ref{sec:Xray}). We caution that selection
criteria used in Ly$\alpha$ surveys to select candidate high redshift
Ly$\alpha$ emitters currently introduce a bias against detecting AGN
with log$[L_B/L_{B,\odot}] \gsim 11.3$ (\S~\ref{sec:broadagn}),
corresponding to the faintest AGN identified in the Chandra Deep
Fields.

{\bf Acknowledgments} Our research is supported by the Australian
Research Council. The authors would like to thank Colin Norman for
helpful discussions and Zolt\'an Haiman for useful comments on an earlier version of the manuscript.

\newcommand{\noopsort}[1]{}

\appendix
\section{Transmission of Ly$\alpha$ Photons from AGN}
\label{app:trans}
For a source of UV radiation embedded in a neutral IGM the emission
blueward of the Ly$\alpha$ line center is suppressed by a factor of
$\langle e^{-\tau_{\rm GP}}\rangle$. Here, the Gunn-Peterson optical
depth, $\tau_{\rm GP}$ is given by
\begin{equation}
\tau_{\rm GP}=\frac{3n(z)A_{\alpha}\lambda_{\alpha}^3}{8\pi H(z)}=6.4 \times 10^5 \Big{(}\frac{\Omega_b h^2}{0.022}\Big{)}\Big{(} \frac{0.15}{\Omega_m h^2}\Big{)}^{\frac{1}{2}}\Big{(}\frac{1+z}{7.5}\Big{)}^{\frac{3}{2}},
\end{equation} 
where $A_{\alpha}=6.25 \times 10^8$ s$^{-1}$ is the Einstein A
coefficient, $\lambda_{\alpha}=1216$ \AA \hs is the Ly$\alpha$
wavelength, and $n(z)$ and $H(z)$ are the number density of neutral
hydrogen atoms and the Hubble constant at redshift $z$, respectively
\citep{Gunn65}. For photons initially redward of the Ly$\alpha$ line
center, the Gunn-Peterson optical depth reduces to
\begin{eqnarray}
\tau_{\rm GP}(x)=\tau_{\rm GP}\frac{1}{\sqrt{\pi}}\int_x^{\infty}\phi(x')dx' \approx-\tau_{\rm GP}\frac{a}{\pi x} \nonumber \\
\approx 10 \Big{(}\frac{1+z}{7.5}\Big{)}^{3/2}
\Big{(}\frac{300{\rm K}}{T_{\rm gas}}\Big{)}^{1/2}\Big{(}\frac{-38}{x}\Big{)}.
\label{eq:taugpx}
\end{eqnarray} 
Here we have expressed the frequency $\nu$ in terms of $x\equiv
(\nu-\nu_0)/\Delta \nu_D$, where $\Delta \nu_D=v_{th}\nu_0/c$, and
$v_{th}=\sqrt{2k_BT/m_p}$ is the thermal velocity of the hydrogen
atoms in the gas, $k_B$ is the Boltzmann constant, $T$ the gas
temperature, $m_p$ the proton mass and $\nu_0=2.47 \times 10^{15}$ Hz
is the central Ly$\alpha$ frequency. Note that $x<0$ for photons
redward of the line center. To obtain the simple expression in
Eq.~(\ref{eq:taugpx}), we approximated the Voigt function $\phi(x)$ in
the line wing as $\phi(x)=a/[\sqrt{\pi}x^2]$, where $a$ is the Voigt
parameter and $a=A_{\alpha}/4 \pi \Delta \nu_D$ $=4.7\times 10^{-4}$ $(13
\hs {\rm km \hs s}^{-1}/v_{th})$ is the ratio of the Doppler to
natural line width.

Another way to write the Gunn Peterson damping wing optical depth is
in terms of a velocity offset, $\Delta$v. For a photon initially
redshifted by $\Delta$v relative to the line center, the total
Gunn-Peterson optical depth reduces to:
\begin{equation}
\tau_{\Delta \rm v}\approx 10 \Big{(}\frac{1+z}{7.5}\Big{)}^{3/2}\Big{(}\frac{85 \hs {\rm km} \hs {\rm s}^{-1}}{\Delta {\rm v}}\Big{)}.
\end{equation} Note that this expression is independent of the gas temperature.

\begin{figure}
\vbox{\centerline{\epsfig{file=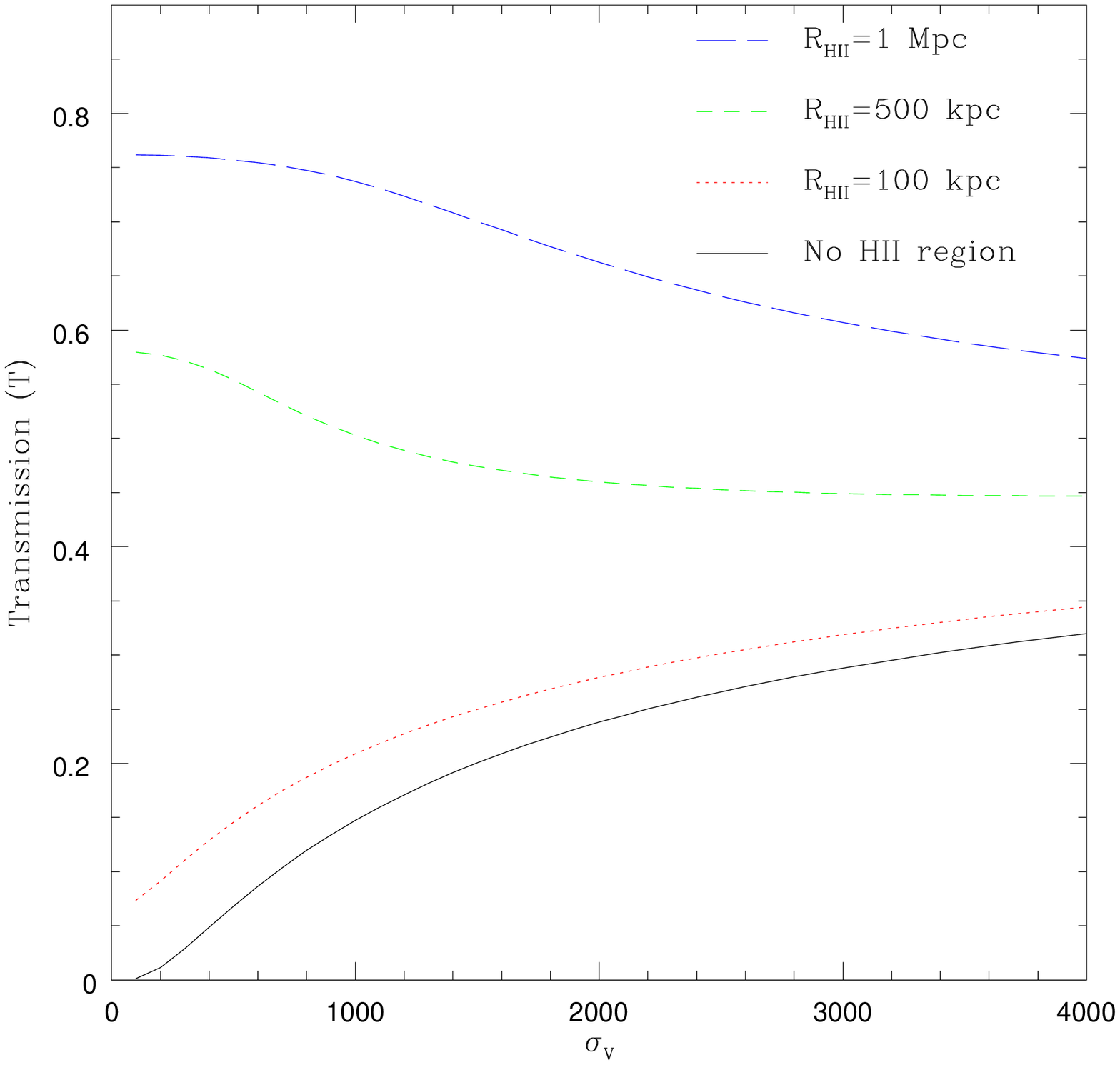,width=8.0truecm}}}
\caption[z=10 Transmission]{The fraction $\mathcal{T}$ of Ly$\alpha$ photons emitted by a $z=6.5$ quasar embedded in a neutral IGM, that is transmitted to the observer, as a function of the intrinsic Ly$\alpha$ line width of the quasar, $\sigma_v=v_{\rm HWHM}$. The four lines represent models with cosmological HII regions of different sizes. The figure shows that an IGM transmission of $\mathcal{T}=0.3$ is accurate to within a factor of 2 for a wide range of parameters.}
\label{fig:transmission} 
\end{figure}

Assuming the intrinsic Ly$\alpha$ spectrum of an AGN to be Gaussian of
width $\sigma_{\rm v}=v_{\rm HWHM}$ (Eq.~\ref{eq:vfwhm}) with a
central flux that is $N$ times the continuum\footnote{This corresponds
to a Ly$\alpha$ EW of $N\times [2v_{\rm HWHM}/c]\times
\lambda_{\alpha}$ \AA=$200 \times (N/10) \times (v_{\rm HWHM}/[3000
\hs{\rm km \hs s}^{-1}])$ \AA.}. The intrinsic flux density
(in arbitrary units) becomes
\begin{equation}
F(x)=1+\frac{N-1}{\sigma_x\sqrt{2 \pi}}{\rm exp}\big{[}\frac{-x^2}{2\sigma_x^2}\big{]},
\label{eq:flux}
\end{equation} where $\sigma_x=\sigma_v/v_{\rm th}$. The IGM transmission is given by

\begin{equation}
\mathcal{T}=\frac{\int_{-\infty}^{\infty}dx [F(x)-1]e^{-\tau(x)}}
{\int_{-\infty}^{\infty}dx [F(x)-1]},
\end{equation} 
where $F(x)$ and $\tau(x)$ are given by Eq.~(\ref{eq:flux}) and
Eq.~(\ref{eq:taugpx}). The transmission $\mathcal{T}$ is plotted as a
function of $\sigma_v$ in Figure~\ref{fig:transmission} as the {\it
black solid line}.

The other lines represent the IGM transmission in the presence of a
cosmological HII region with a radius of $0.1$ ({\it red--dotted
line}), $0.5$ ({\it green--dashed line}) and 1.0 Mpc ({\it
blue--long--dashed dotted line}) surrounding the AGN. The radius of
the HII region surrounding a quasar of age $t_Q$ powered by a black
hole of mass $M_{\rm BH}$ shining at Eddington luminosity is given by
\citep{CenHaiman02}
\begin{equation}
R_{\rm HII, edd}=0.74 \Big{(}\frac{M_{\rm BH}}{10^7 M_{\odot}} \Big{)}^{1/3}\Big{(}\frac{t_Q}{2 \times 10^7{\rm yr}} \Big{)}^{1/3}\Big{(} \frac{1+z}{7.5}\Big{)}\hs {\rm Mpc}.
\end{equation} 
This relation assumes the total ionizing luminosity of AGN, $L_{\rm
ion}$, to scale as $L_{\rm ion}\propto M$. Alternatively, the Kaspi
relation shows that the continuum luminosity at $\lambda=5100$ \AA \hs
scales as $ \propto M^{1.3}$ (Eq.~\ref{eq:kaspi}). If the brightest
$z>6$ AGN accrete at the Eddington luminosity, and we assume that the
total ionizing luminosity emitted by an AGN also scales as $M^{1.3}$,
we get
\begin{equation}
R_{\rm HII, kas}=0.44 \Big{(}\frac{M_{\rm BH}}{10^7 M_{\odot}} \Big{)}^{0.43}\Big{(}\frac{t_Q}{2 \times 10^7{\rm yr}} \Big{)}^{1/3}\Big{(} \frac{1+z}{7.5}\Big{)}\hs {\rm Mpc}.
\end{equation}

The size of the HII region is also determined by the uncertain value
of the escape fraction of ionizing photons, $f_{\rm esc}$. Accounting
for these uncertainties, Figure~\ref{fig:transmission} shows that for
$\sigma_v=3000$ km s$^{-1}$, $\mathcal{T}$ lies in the range
$0.3-0.45$ given $R_{HII}=0-500$ kpc. The range of
$\mathcal{T}$ increases towards lower $\sigma_v$. However we find that
an IGM transmission of $\mathcal{T}=0.3$ is accurate to within a
factor of 2 for a wide range of the parameters $\sigma_v$, $f_{\rm
esc}$ and total ionizing luminosity.

\begin{figure}
\vbox{ \centerline{\epsfig{file=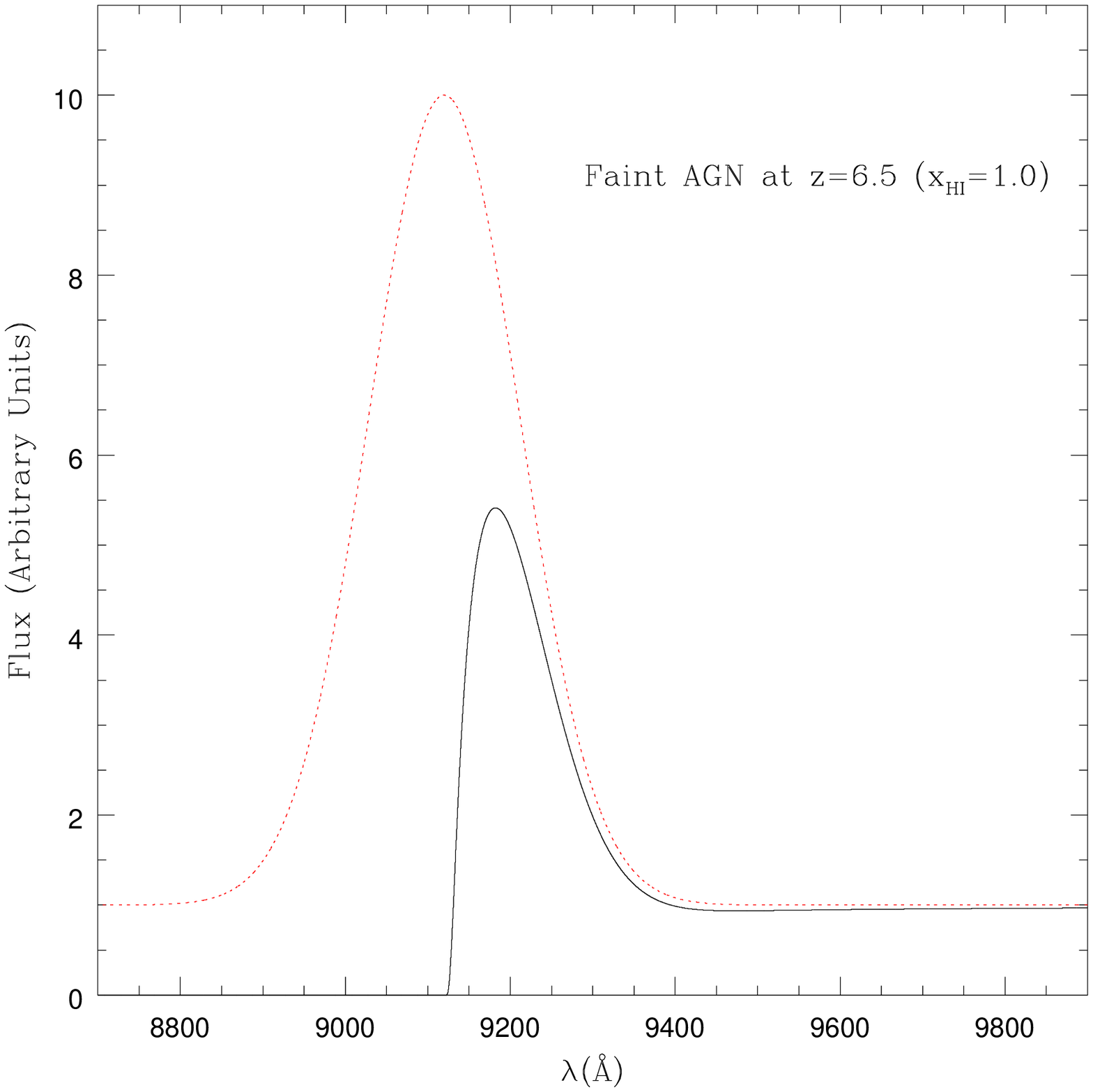,width=8.0truecm}}}
\caption[z=10 QSO]{The observed spectrum of a faint AGN at $z=6.5$ in a neutral IGM (i.e. no cosmological HII region is present) is shown as the {\it black solid line}. The intrinsic Ly$\alpha$ line is shown as the {\it red dotted line}. A neutral IGM erases a large fraction of the line, transmitting only the reddest photons to an observer. The figure shows that these AGN have a peak Ly$\alpha$ flux redward of the true line center by $\sim 60$ \AA (corresponding to $\sim 2000$ km s$^{-1}$) and an observed $v_{\rm HWHM}$ that is a factor of $\sim 2$ times lower than the intrinsic $v_{\rm HWHM}$.}
\label{fig:qso} 
\end{figure}

In Figure~\ref{fig:qso} the {\it black solid line} shows the
theoretical spectrum of a $z=6.5$ AGN embedded in a neutral IGM
(i.e. no cosmological HII region is present). We assume the black hole
mass to be $10^7M_{\odot}$, which according to Eq.~(\ref{eq:vfwhm})
should yield $v_{\rm HWHM}=4 \times 10^{3}$ km s$^{-1}$. However, at
low luminosities, and therefore low black hole masses, $v_{\rm
HWHM}\rightarrow 3\times 10^{3}$ km s$^{-1}$ \citep{Kaspi00}. The
intrinsic Ly$\alpha$ spectrum (the {\it red dotted line}) is
described by Eq.~(\ref{eq:flux}), in which we assumed
$\sigma_v=3\times 10^3$ km s$^{-1}$ and $N=10$ (which yields an
Ly$\alpha$ EW of 200 \AA).

 We note that the observed line center lies redward of the true line center by $\sim 60$ \AA, which translates to $\sim 2000$ km s$^{-1}$. Furthermore, the observed $v_{\rm HWHM}$ is $\sim 2$ times lower than the emitted value.
 
\label{lastpage}
\end{document}